\documentclass[aps,prd,amsmath,amssymb,preprintnumbers,preprint,nofootinbib,a4paper,11pt]{revtex4-1}
\pdfoutput=1
\usepackage{graphicx}
\graphicspath{{./figures/}}
\usepackage{url}
\usepackage{color}
	\definecolor{rossoCP3}{cmyk}{0,.88,.77,.40}
		\definecolor{graa}{rgb}{0.8,0.8,0.8}
		\definecolor{blaa}{rgb}{0.2,0.2,0.6}
\usepackage{bm}
\usepackage{bbm}
\usepackage{pxfonts}
\usepackage[margin=5pt, font=small,labelfont=bf,justification=raggedright]{caption}
\usepackage{subfig}
\usepackage{youngtab}
\usepackage{slashed}
\usepackage{cancel}

\newcommand{\beq}{\begin{eqnarray}}
\newcommand{\eeq}{\end{eqnarray}}

\newcommand{\ea}[1]{
\begin{align}
#1
\end{align}
}

\newcommand{\bmp}{\noindent\begin{minipage}{16cm}}
\newcommand{\emp}{\end{minipage}\vskip 7mm} 

\newcommand{\Tr}{\text{Tr}}

\newcommand{\Eref}[1]{\eqref{#1}}

\newcommand{\mean}[1]{\langle #1 \rangle}

\begin{document}
\phantom{g}\vspace{2mm}
\title{ \LARGE  \color{rossoCP3} 
Jumping Out of the Light-Higgs  Conformal Window} \author{Oleg {\sc Antipin} 
}\email{antipin@cp3-origins.net}
\author{Matin {\sc Mojaza}
}\email{mojaza@cp3-origins.net} 
\author{Francesco {\sc Sannino}
}\email{sannino@cp3.dias.sdu.dk} 
\affiliation{
{ \color{rossoCP3}  \rm CP}$^{\color{rossoCP3} \bf 3}${\color{rossoCP3}\rm-Origins} \& the Danish Institute for Advanced Study {\color{rossoCP3} \rm DIAS},\ \\ 
University of Southern Denmark, Campusvej 55, DK-5230 Odense M, Denmark.
}
\begin{abstract}
We investigate generic properties of the conformal phase transition in gauge theories featuring Higgs-like fundamental particles. These theories provide an excellent arena to properly investigate conformal dynamics and discover novel features. 
We show that the phase transition at the 
boundary of the Higgs conformal window is not smooth but a jumping one for the known perturbative examples. In addition the general conditions under which the transition is either jumping or smooth are provided.
Jumping implies that the massive spectrum of the theory will jump at the phase transition. It, however, still allows for one of the states, the would be dilaton of the theory, to be lighter than the heaviest states in the broken phase. Finally we exhibit a calculable Higgs model in which we can, in perturbation theory, determine the Higgs conformal window, the spectrum in the conformally broken phase and demonstrate it to possess a jumping type conformal phase transition. 
  \\[.1cm]
{\footnotesize  \it Preprint: CP$^3$-Origins-2012-21 \& DIAS-2012-22}
\end{abstract}
\maketitle
\tableofcontents

\newpage
\section{Introduction}

{The recent experimental discovery, claimed by the Large Hadron Collider experiment at CERN, of a new boson with mass around $126$~GeV must be taken into account in any extension of the standard model of particle interactions \cite{:2012gu,:2012gk,:2012zzl}.  It is a fact that the new state, be it elementary or composite\cite{Frandsen:2012rj}, is light with respect to the natural scale of the electroweak theory which is   $4\pi v\simeq 3.1$~TeV with $v\simeq 246$~GeV.  It is therefore very topical to set up and investigate  generic properties of the (near) conformal dynamics associated to gauge theories involving fermions and fundamental scalars mimicking the Higgs. }

\begin{figure}[h]
\centering
\includegraphics[width=0.4\columnwidth]{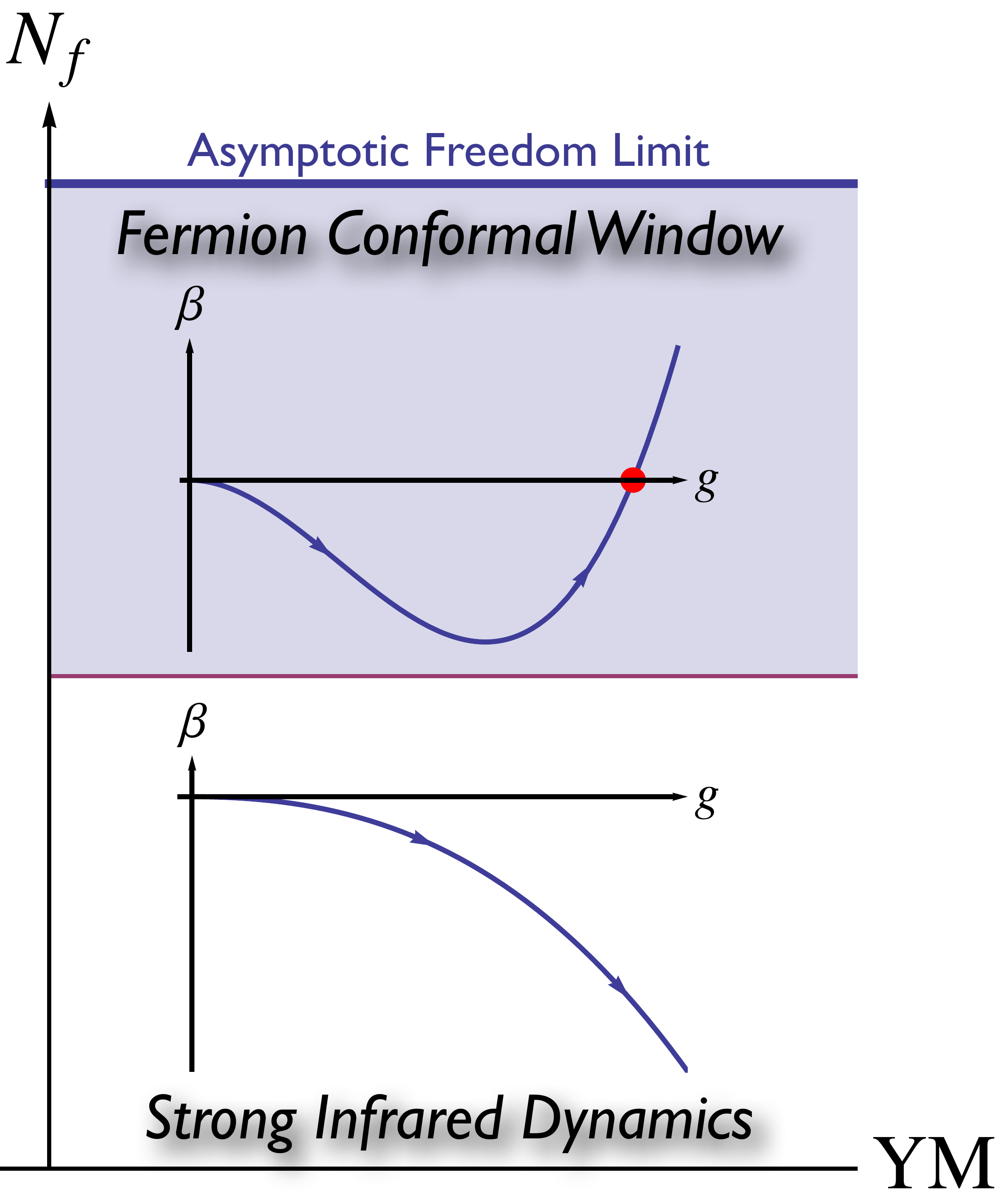} 
 \caption{Fermion conformal window of a gauge theory with $N_f$ fermionic matter in a generic representation of a gauge group. The running behavior of the gauge coupling is illustrated by the dependence of the beta function, $\beta = dg/d\ln\mu$, on the gauge coupling in the two regions. Note that here we are not speculating  on the behavior of the coupling near the lower boundary of the conformal window.}
\label{FCW}
\end{figure}

Recent studies of gauge theories including both fermion and scalar fields and containing a perturbative infrared stable fixed point in the renormalization group flow have led to new developments in the understanding of the loss of large-distance conformality \cite{Grinstein:2011dq,Antipin:2011aa,Antipin:2012kc,Antipin:2011ny}.
In particular, it has been shown that such theories have at least  two types of conformal windows
\footnote{The {\it conformal window} is the term used to denote the region in parameter space where the theories do not develop an intrinsic scale, as in QCD, but where the couplings run to an infrared stable fixed point and the theories freeze in to a conformal field theory.}.
{For example}, in gauge theories with purely fermionic matter, the conformal window spans a region in the \emph{external} parameter space being the number of fermionic flavors of each representation of the given underlying gauge group. Below the conformal window where the theories are still asymptotically free an intrinsic scale is believed to develop due to condensation of the fermions. This is illustrated in Fig. \ref{FCW}.

\begin{figure}[h]
\centering
\includegraphics[width=0.75\columnwidth]{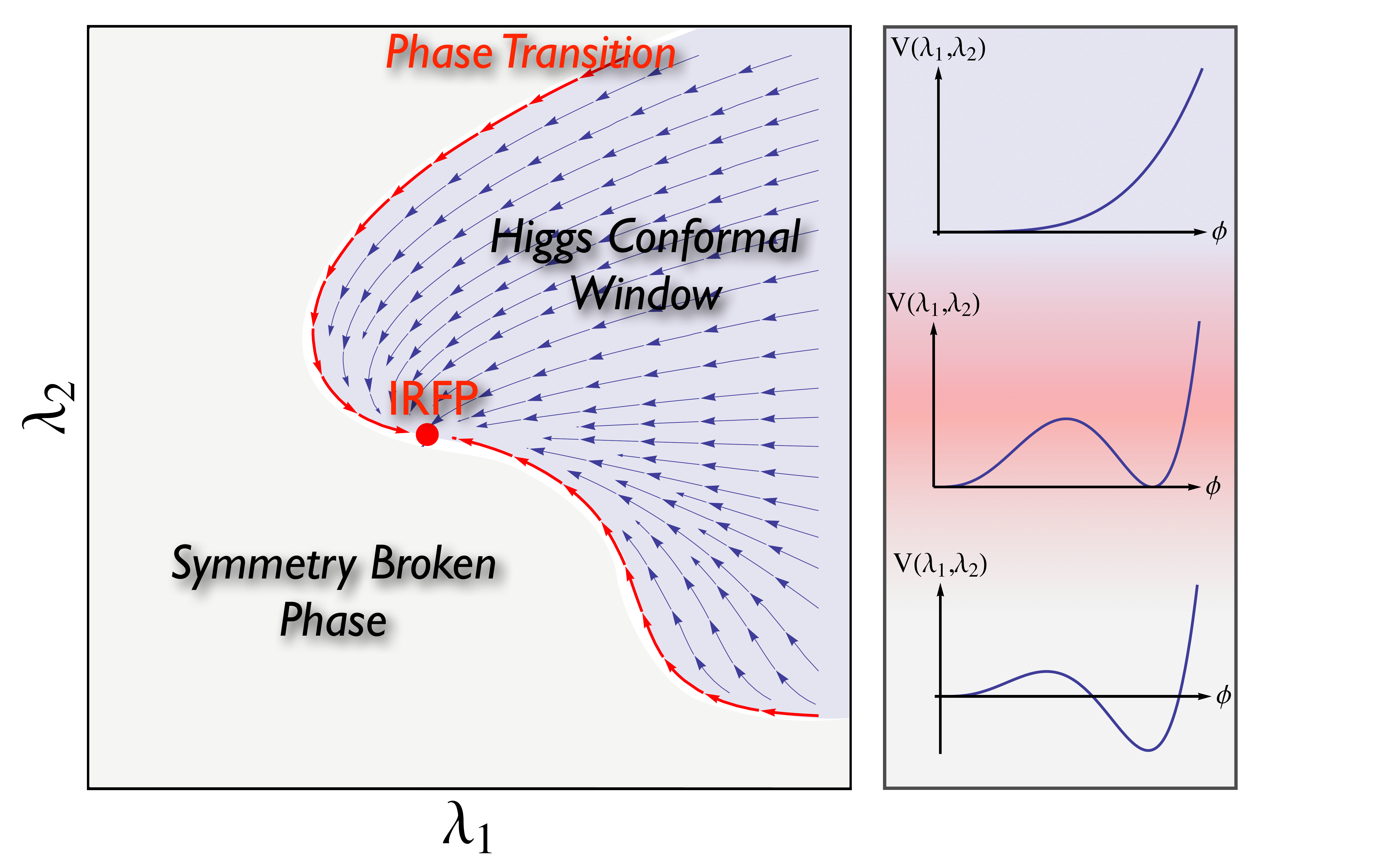} 
 \caption{Higgs conformal window of a theory with bosonic matter and an infrared stable fixed point (IRFP) in the renormalization group flow. $\lambda_1$ and $\lambda_2$ represent some running couplings, upon which the effective scalar potential depends. In the symmetry broken phase, the couplings are decoupled at the symmetry breaking scale and therefore do not run. }
\label{SCW}
\end{figure}

{Another} possible kind of conformal window arises in gauge theories including bosonic matter. This extent of the window is due to the possibility of developing a scale not by fermion condensation, but by the Higgs acquiring a non-trivial vacuum expectation value (vev). This type of conformal window is necessarily a span in the \emph{internal} parameters, i.e. the coupling constants of the theory. Of course, to have large-distance conformality the Higgs quadratic terms in such theories are fine-tuned to zero and the conformal window is a span in \emph{dimensionless} coupling constants only. The border of the Higgs conformal window is marked by the onset of the Coleman-Weinberg phenomenon \cite{Coleman:1973jx} that induces a non-trivial vev by radiative corrections to the effective scalar potential.
This is illustrated in Fig. \ref{SCW}, where the phase transition is assumed to be jumping \cite{Sannino:2012wy}, i.e. of the first order.

In explicit perturbative model examples  \cite{Grinstein:2011dq, Antipin:2011aa} it was shown that in the crossing out of the Higgs conformal window, a dilatonic state appears with mass parametrically lighter than the heaviest states and couples strongest to the dilatation current due to the existence of the underlying IR fixed point (IRFP) in the renormalization group (RG) flow. It was {argued } in \cite{Grinstein:2011dq} that this feature is due to an underlying walking-type dynamics \cite{Holdom:1981rm,Holdom:1984sk,Yamawaki:1985zg,Appelquist:1986an} of the theory near the conformal boundary. This requires inherently that the phase transition across the conformal phase boundary is of second order or higher. This statement was made using the unimproved effective potential of the theory which, in fact, fails to provide accurate results near the origin of the Higgs field.
By the use of the improved effective Higgs potential technique, which allows to extend the validity of the effective potential to the origin of the Higgs field, we will show that this is not necessarily the case in general. In fact at the cross-over from the Higgs conformal window to the broken side the transition in the known perturbative examples \cite{Grinstein:2011dq, Antipin:2011aa} is actually of first order.  Although this inevitably implies a jump of the entire spectrum of the theory at the cross over it still allows for large mass scale separations between the states of the theory when jumping out of the conformal window.  {We will also argue that the dilaton can still be lighter than the heaviest states in the theory.}

In section \ref{Trace} we introduce the relevant notation and set up the stage by discussing how to correctly improve the Higgs potential at any order in perturbation theory. We then investigate general consequences for theories featuring nontrivial FPs. We discuss how to properly define the boundary of the Higgs conformal window and how to determine whether the conformal phase transition is of walking or jumping type. We provide an explicit calculable example in section \ref{AMS} and conclude in section \ref{conclusions}.

\section{Trace Anomaly, Higgs Conformal Window and the Dilaton}
\label{Trace}
In quantum field theories that are classically conformally invariant,
the conformal symmetry is generally explicitly broken by the trace anomaly.
This is so, since the trace of the improved energy-momentum tensor $\Theta_{\mu \nu}$, 
equals the divergence of the dilatation current $D^\mu$:
\begin{align}
\partial_\mu D^\mu = \Theta_\mu^\mu .
\end{align}
The trace of $\Theta_{\mu \nu}$ is given by
\begin{align}
 \Theta_\mu^\mu = \sum_g \beta(g) \frac{\partial \mathcal{L}}{\partial g} \ ,
\end{align}
where $g$ collectively denotes all dimensionless coupling constants of the theory,
$\beta(g)$ is the beta function defined as $dg/d\ln \mu = \beta(g)$ and $\mathcal{L}$ is the Lagrangian density.
Since this trace is generally non-zero, scale invariance, or dilatation symmetry, of the classical theory is anomalous and is therefore explicitly broken by this \emph{trace anomaly}.
An obvious exception to this rule is if the coupling constants $g$ take a fixed point value $g^*$ such that
\begin{align}
\beta(g^*) = 0 \ ,
\end{align}
for all couplings.
If a classically scale invariant theory has an IRFP, the scale invariance will be fully restored at the quantum level in the deep infrared if the fixed point is reached. The fixed points can, however, {disappear} if scale invariance is broken \emph{spontaneously} at an intermediate energy scale between the UV and the IR. This can happen either by strong dynamics inducing condensation of certain operators,  or by the Higgs potential developing a global minimum away from the origin due to the running of the couplings. {Of course, in general, both phenomena can emerge simultaneously. Because the strongly interacting regime poses a formidable challenge we start by shedding light on some of these phenomena by investing the Higgs condensation phenomenon induced by the Coleman-Weinberg mechanism \cite{Coleman:1973jx} near perturbative fixed points.}
Recent investigations of this type \cite{Grinstein:2011dq, Antipin:2011aa}, in models where all computations are under perturbative control, 
reveal interesting features of the low-energy spectrum outside the Higgs conformal window. In particular, it is possible to unambiguously identify a dilaton in the spectrum whose mass is vanishing as one approaches the phase separating boundary (separatrix) of the conformal window. Moreover, it turns out that the dilaton is not necessarily the lightest particle state of the spectrum as one moves away from the separatrix.

\subsection{Setting up the stage and identifying the caveats}
\label{naiveSCW}
{We now extract some generic features of the Higgs conformal window, starting from 
 what has been learned in the two model examples investigated so far \cite{Grinstein:2011dq, Antipin:2011aa}.}
The prerequisite for a theory to have a Higgs conformal window is:
\begin{center}
\it A theory containing fundamental Higgs fields with an infrared stable fixed point.
\end{center}
The renormalized mass of the scalars is, from the outset fine-tuned to be zero  \cite{Grinstein:2011dq, Antipin:2011aa}. Thus, the Higgs potential is given solely in terms of dimensionless couplings.
For the actual existence of a Higgs conformal window in such theories,
is required that
\begin{center}
\it The coupling constants space must be partly unstable against the Coleman-Weinberg phenomenon.
\end{center}
This phenomenon will induce a scale by dimensional transmutation and thus
prevent the theory from displaying large-distance conformality, or in RG terms, {modify the running of the couplings in such a way that they cannot reach the fixed point}. 
The Coleman-Weinberg (CW) phenomenon happens if 
a Higgs develops a vacuum expectation value different from zero.
The vacuum can be studied by investigating the CW effective potential
which is given in terms of {classical} Higgs fields, and
the minima of the effective potential gives, {in perturbation theory}, the vacuum state of the theory.

Let us consider this in greater detail and look at how it applies to theories with an IRFP.  Consider a classically conformally invariant theory with one real scalar field $\phi$ coupled to fermions and/or gauge fields. Denote the coupling constants by $g$, collectively.
The classical field $\phi_c$ is defined through the generating functional formalism, where one adds a source term $J(x)\phi(x)$ to the Lagrangian density, such that:
\begin{align}
\phi_c(x) = \frac{\delta W}{\delta J(x)} = \frac{\langle 0^+ | \ \phi(x) \ | 0^-\rangle}{\langle 0^+ | 0^-\rangle}\Big |_J \ ,
\end{align}
where $W$ is the generating functional and $| 0^\pm \rangle$ are the asymptotic vacuum states of the far past and future.
The effective potential takes the general form:
\ea{
V (\phi_c ) = \phi_c^4 F\left (t, g \right) \ ,
}
where for dimensional reasons, $F$ can only be a function of dimensionless parameters, i.e. the coupling constants, $g$, and the ratio $\phi_c/\mu$, which we parametrize logarithmically by
\ea{
t \equiv \ln \frac{\phi_c}{\mu} \ ,
}
and $\mu$ is the renormalization scale at which the set of coupling constants $g$ is defined.
$F$ is some function which,  {when possible, can be determined perturbatively}:
\ea{
\label{Fbare}
F\left (t, g \right) = A(g) + B(g) t + C(g) t^2 + \cdots \ .
}
{Imposing perturbative reliability, we require} that $|g|\ll 1$ and $|g t | \ll 1$.
To leading order in $t$, the effective potential has a minimum at
\ea{
\label{tsbare}
t^* = \ln \frac{\mean{\phi_c}}{\mu}=-\frac{1}{4} - \frac{A(g)}{B(g)} \ ,
}
if 
\ea{
V^{\prime\prime} (\mean{\phi_c}) = 4 B(g) \mean{\phi_c}^2 > 0
}
{This local minimum can be trusted} if $|gt^*| \ll 1$, that is if $A(g) \sim B(g)$ are of the same order. {To set the stage it is useful to first summarize the naive analysis provided above and then discuss the associated caveats and how to perform the actual, proper analysis:}
\begin{itemize}
\item
 The region in the coupling constant space where $B(g) > 0$ is unstable against the Coleman-Weinberg phenomenon. In RG terms, any RG trajectory passing through this region will not reach the IRFP, but will undergo spontaneous symmetry breaking at the scale $\mean{\phi_c}$. The would-be IRFP therefore cannot be reached.

\item
 The region in coupling constant space where $B(g)<0$ does not, to this order of perturbation theory, lead to symmetry breaking. In RG terms, all trajectories passing through this region will reach the IRFP. This region is the naive perturbative {\emph{Higgs conformal window}}.

\item
 The RG flow through the region in coupling constant space where $B(g) = 0$ defines the {\it separatrix}, i.e. the RG trajectories defining the boundary between the conformal phase from the broken one. Approaching the separatrix from the broken side leads to the (misleading) result $\mean{\phi_c} \to 0$ smoothly.
\end{itemize}
The first caveat is that the analysis above is perturbatively reliable if also the IRFP is accessible in perturbation theory.  This condition helps with investigating the values of $\phi_c$ close to the origin.  The second caveat is that the naive boundary between the two phases introduced above is ill defined since for $B(g) \to 0$ one has $| t^*| \to  \infty$, implying that near the boundary the simple analysis above does not lead to a perturbative reliable minimum. Another way to see this is that the potential, at the origin $\phi_c=0$, has turned into a maximum due to the logarithmic singularity. These problems were already discussed in the original paper by S. Coleman and E. Weinberg \cite{Coleman:1973jx} and will be overcome below. However the qualitative features itemized above still provide a first rough understanding of the phenomenon.

\subsection{Improving by Resumming the Logs} 
The problem of $|t^*|\to \infty$ can be cured using the RG by recognizing that the effective potential must satisfy the RG equation:
\ea{
\frac{d V(\phi_c)}{d\ln \mu} = \left [
\mu \frac{\partial}{\partial \mu} + \sum_g \beta(g)  \frac{\partial}{\partial g} + \gamma(g) \phi_c  \frac{\partial}{\partial \phi_c} 
\right ] V(\phi_c) = 0 \ ,
}
where $\gamma(g)$ is the anomalous dimension of the Higgs field.
This leads to the RG equation for $F$:
\ea{
\label{RGF}
\left [
-  \frac{\partial}{\partial t} + \sum_g \bar{\beta}(g)  \frac{\partial}{\partial g} + 4 \bar{\gamma}(g) 
\right ] F(t, g) = 0 \ ,
}
where
\ea{
\bar{\beta}(g) = \frac{\beta(g)}{1-\gamma(g)}, \qquad \qquad \bar{\gamma}(g) = \frac{\gamma(g)}{1-\gamma(g)} \ .
}
It is important to stress that a fixed point $g^*$ in $\beta$ is also a fixed point of $\bar{\beta}$:
\ea{
\label{ImprovedFP}
\beta(g^*) = 0 \quad \Rightarrow \quad \bar{\beta}(g^*) = 0 \ ,
}
except for the particular case where $\gamma(g^*) = 1$. Unless otherwise stated we assume that this particular case does not happen.  The RG equation for $F$ allows one to substitute away the dependence on one of its variables. This is exactly what is needed to cure the $|t^*| \to \infty$ problem, i.e. we simply improve the perturbative expansion of $F$ by substituting its explicit $t$ dependence with the aid of its RG equation. This procedure corresponds to \emph{resumming the logs} and goes as follows:

By defining new effective running couplings, $\bar{g}$ through the differential equations:
\ea{
\label{barg}
\frac{ d \bar{g}}{d t} = \bar{\beta}(\bar{g}) \ , \quad \text{with} \quad 
\bar{g}(0) = g ,
}
a general solution of \Eref{RGF} can be written in terms of these couplings, as follows:
\ea{
\label{Fimproved}
F(t, g) = f \left( \bar{g}(t)\right ) e^{4 \int_0^t \bar{\gamma}\left( \bar{g}(s) \right)d s}  \ ,
}
where $f$ is an arbitrary functional of the functions $\bar{g}$. It can only be fixed through a matching with the perturbative expansion of $F$. Let us show how this is done to leading order.
 
If we let $\lambda$ be the quartic self-coupling of $\phi$, i.e.
\ea{
V_0 =  \lambda \phi^4 ,
}
then to zeroth order, for $t=0$, $f$ must satisfy :
\ea{
f \left( \bar{g}(0)\right ) = \lambda + \mathcal{O}(g^2)  = \bar{\lambda}(0) +  \mathcal{O}(g^2) \ .
}
This leads to the unique lowest-order expansion in terms of $\bar{g}_i = \{\bar{\lambda}, \bar{g}_2, \bar{g}_3, \ldots \}$ :
\ea{
\label{leadinglog}
f \left( \bar{g}(t) \right ) = \bar{\lambda}(t) + \cdots 
}
By the remaining couplings $g_2, g_3, \ldots$ we actually mean Yukawa and gauge couplings \emph{squared} as they only enter as such in loop expansions. 
Thus requiring $g_i \ll 1$ implies for a Yukawa coupling $y$ that $y^2 \ll 1$. We stress that $\lambda$ and $y^2$ are  to be considered of the same order in perturbation theory.

Using that $\bar{g}(t) = g + \bar{\beta}(\bar{g}) t + \mathcal{O}(t^2)$, we get the leading expansion in $g$ and $t$:
\ea{
f \left( \bar{g}(t) \right )  = \lambda + \bar{\beta}(\lambda) t + \mathcal{O}(g^2, t^2) \ .
} 
Expanding also the exponential factor:
\ea{
 \exp\left({4 \int_0^t \bar{\gamma}\left( \bar{g}(s) \right)d s}\right) 
 \approx \exp\left({4 \int_0^t \sum_{i} \gamma_{i} g_i d s}\right) = 1+ 4 \sum_{i} \gamma_{i} g_i t + \mathcal{O}(g^2 t, t^2) \ ,
 }
 leads to the unique matching of the expansion of $F$ in \Eref{Fimproved} and \Eref{Fbare} leaving no free coefficients to fix for $f$ since $\gamma_i$ can be computed in perturbation theory:
\ea{
\lambda +  \left ( \bar{\beta}(\lambda) + 4 \lambda \sum_i \gamma_i g_i\right ) t 
= A + B t \ .
 }
 To leading order in $g$ we simply have $A= \lambda + \mathcal{O}(g^2)$.
 Thus
 \ea{
 \label{B}
 B = \beta(\lambda) + 4 \lambda \sum_i \gamma_i g_i + \mathcal{O}(g^3) \ .
 }
 The bar on $\beta$ is removed, since to leading order in $g$, $\bar{\beta} = \beta$.
We note that this identity must always hold, i.e. $B$ can either be explicitly computed from perturbation theory, or is given by the above expression once the one-loop beta function and $\gamma$ are known. The solution \Eref{leadinglog} thus recovers the one-loop leading-log result, which is of $\mathcal{O}(g^2 t)$, but not the one loop terms in $A$, which are of $\mathcal{O}(g^2 \ln g)$ and $\mathcal{O}(g^2)$. Solution \eqref{leadinglog} is called the leading log (or tree-level) improved effective potential. The procedure outlined above of improving the effective potential can be performed order by order in perturbation theory. For the benefit of the reader we summarize it below:
{\begin{enumerate}
\item 
By knowing the one-loop beta functions and anomalous dimension, 
the tree-level improved effective potential is directly given by:
\ea{
V_{\rm tree}(\phi_c) = A^{(0)}\left(\bar{\lambda}(t)\right) \phi_c^4 \ e^{4 \int_0^t \bar{\gamma}\left( \bar{g}(s) \right)d s} ,
}
where $A^{(0)}\phi_c^4$ is the zeroth-order effective potential with $\lambda$ replaced by $\bar{\lambda}(t)$, given by \eqref{barg}. All leading logs are contained in this potential term, as demonstrated above.
\item
Knowing $A^{(1)}$ of the one-loop effective potential and knowing the \emph{two-loop} beta functions and anomalous dimension, the one-loop improved effective potential is given by:
\ea{
V_{\rm 1loop}(\phi_c) = \left (A^{(0)}\left(\bar{\lambda}(t)\right) + A^{(1)}\left(\bar{g}(t)\right) \right )\phi_c^4 \ e^{4\int_0^t \bar{\gamma}\left( \bar{g}(s) \right)d s} .
}
All leading logs and next-to-leading logs are then contained in this expression. Note that $A^{(1)}$ is in general a function of all the couplings $g$.
\item
The generic structure of improving the effective potential to any order should be clear from the above. To $L$-th order, it reads:
\ea{
\label{improvedpot}
V_{ L\rm-loop}(\phi_c) = A\left(\bar{g}(t)\right) \phi_c^4 \ e^{4 \int_0^t \bar{\gamma}\left( \bar{g}(s) \right)d s} \ ,
}
where $A$ is evaluated to $L$ loops:
\ea{
A\left(\bar{g}(t)\right)
= A^{(0)}\left(\bar{\lambda}(t)\right) + \cdots + A^{(L)}\left(\bar{g}(t)\right) \ .
}
Hence, the function $f$ is nothing but $A$ computed in perturbation theory with the bare couplings $g$ replaced by the improved effective couplings $\bar{g}$.
 It is important to note that one needs the beta functions and $\gamma$ to ($L+1$)-loop order  to improve the $L$-loop effective potential.
\end{enumerate}
}
In practice, we have managed to i) hide away the explicit $t$ dependence of the effective potential and ii) take the running of couplings into account when surveying the values of $\phi_c$ away from $\mu$. The validity of perturbation theory now only requires $\bar{g}(t) \ll 1$, since the procedure above makes sure that $t$, or the associated logs in the effective potential, can become large without jeopardizing the perturbative expansion. This is a quite remarkable result especially, in the context of infrared or ultraviolet fixed points as it will be clear in the following section.

\subsection{Improved Potentials for Theories with Fixed Points}
\label{potential}
It follows immediately from \Eref{ImprovedFP} that in the presence of an infrared (ultraviolet) (IR) stable
fixed point in the running couplings we must have 
for $\phi_c \to 0$ ($\phi_c \to \infty$) that $\bar{g} \to g^*_{\rm IR} $ ($\bar{g} \to g^*_{\rm UV} $) independently of the bare coupling constant values\footnote{Still, of course, within the correct basin of attraction of the fixed point. }, i.e.,
\begin{equation}
\lim_{t\to -\infty } \bar{g}(t) = g^*_{\rm IR}  \quad 
( \lim_{t\to \infty } \bar{g}(t) = g^*_{\rm UV} ) \ ,\quad 
\text{for any} \quad \bar{g}(0) = g   \ .
\end{equation}

 Since the couplings runs very slowly near a fixed point, we can 
 make a few rigorous statements about the
 asymptotic form of the effective potential.
 
When having a stable IRFP, then near $\phi_c = 0$ the improved effective potential is well approximated by
\ea{
V (\delta \phi_c ) = A(g^*_{\rm IR}) \ \delta\phi_c^4 \ e^{-4 \int_{-\infty}^0 \bar{\gamma}\left( \bar{g}(s) \right)d s} \ ,
}
where $\delta \phi_c$ is a small perturbation away from zero, with small being $\delta \phi_c \ll \mu$. This result holds for \emph{any} $\mu$, once there is an IR stable fixed point. This tells us that the curvature of the effective potential at the origin is 
the same for \emph{any} chosen values of the bare couplings and is solely 
determined by the sign of $A$ at the fixed point.
Thus, 

\begin{center}
\it
In a theory containing scalars with an IR stable fixed point the curvature of the Higgs potential at the origin for any values of the theory parameters, is solely determined by the potential computed at the fixed point.
\end{center}
{The statement above needs to be made since one has to be able to properly follow the RG trajectories untill the fixed points.}
Another interesting statement can be made on the boundedness of the effective potential in asymptotically safe theories, i.e. ones that contain an UV stable FP at $g_{\rm UV}^*$. In this case for any value of the bare couplings, connected to the UVFP by a RG flow, the relevant sign of the potential is again solely controlled by the value at the fixed point:
\begin{equation}
\lim_{\phi_c \to \infty} \frac{V(\phi_c)}{\phi_c^4} = A(g_{\rm UV}^*) \ e^{4 \int^{\infty}_0 \bar{\gamma}\left( \bar{g}(s) \right)d s} \ .
\end{equation}
In particular, to have a potential bounded from below, the fixed point must satisfy $A(g_{\rm UV}^*) \geq 0$. In asymptotically free theories, where $g_{\rm UV}^*=0$, the inequality is exactly saturated, since $A(g_{\rm UV}^*)=0$. Thus,

\begin{center}
\it
In a theory containing scalars with an UV fixed point the scalar potential
is bounded for any values of the theory parameters, if the effective quartic coupling at the fixed point is positive.
\end{center}

\subsection{Resolving the Boundary of the Higgs Conformal Window}
\label{Boundary}
Having resolved the issue with $|t^* |\to \infty$ of section \ref{naiveSCW} by the aid of the RG we are now able
to precisely define the boundary of the Higgs conformal window  in perturbation theory. This can be readily done by searching for a non-trivial minimum of the improved effective potential as given in \Eref{improvedpot}:
\ea{
\label{diff}
0 &= \frac{d V}{d\phi_c} = \left[{4 A(\bar{g}(t^*))+ \frac{\partial A}{\partial \bar{g}_i}\beta(\bar{g_i}(t^*))} \right]  \frac{\mean{\phi_c}^3 \ e^{4 \int^{t^*}_0 \bar{\gamma}\left( \bar{g}(s) \right)d s}}{1-\gamma(\bar{g}(t^*))}  \ , \\[2mm]
\label{ddiff}
0& < \frac{d^2 V(\mean{\phi_c})}{d\phi_c^2}  = 
\left [ \frac{\partial A}{\partial \bar{g}_i} \left(4 + \beta(\bar{g}_j) \frac{\partial}{\partial \bar{g}_j}\right )+  \beta(\bar{g}_j) \frac{\partial^2 A}{\partial \bar{g}_j \partial \bar{g}_i} \right] \beta(\bar{g_i})  
\frac{\mean{\phi_c}^2 \ e^{4 \int^{t^*}_0 \bar{\gamma}\left( \bar{g}(s) \right)d s}}{\left (1-\gamma(\bar{g}(t^*))\right)^2}  \ .
}
where summation over repeated indices is assumed. Furthermore, if the potential at such minimum is positive, it can only be a local one, since $V(0) = 0$. Thus, we also want
\ea{
V(\mean{\phi_c}) < 0 \quad \Rightarrow \quad A(\bar{g}(t^*)) < 0 \ .
}
If such $t^* \neq - \infty$ exists for some coupling constant values $g$, spontaneous symmetry breaking will occur before reaching the IR fixed point. The region of coupling constant space where there is no spontaneous chiral symmetry breaking defines the Higgs conformal window. 
The separation between the two regions, i.e. the boundary of the conformal window, is now well defined by the condition:
\ea{
\label{SCWboundary}
A(\bar{g}(t^*)) = 0,
} 
for a $t^* \neq - \infty$ and satisfying \Eref{diff} and \Eref{ddiff}.

It is instructive to see how it relates to the naive definition from the bare potential analysis in section \ref{naiveSCW}. At one loop, we have that $A(\bar{g}) = \bar{\lambda}(t)$.
This holds for any $t$ as long as $\bar{\lambda} (t) \ll 1$.
The bare definition, on the other hand, is only valid for $t\lesssim 1$.
To compare, we must therefore restrict to the case where  $t^*$ is small and
make a perturbative expansion of $\bar{\lambda}$ in $t^*$, which reads:
\ea{
0 = \bar{\lambda}(t^*) \approx \lambda + \beta(\lambda) t^*
}
The comparison of the {\it improved} condition above with the one loop bare result $B=0$, with the unimproved $B$ given in \Eref{B},
then reads: 
\ea{
\text{Bare:} \quad \beta(\lambda) + 4\gamma_i g_i \lambda = 0, \qquad \text{Improved}: \quad \beta(\lambda) + \frac{\lambda}{t^*} = 0 \ .
}
While the functional form of the two definitions match up, they cannot coincide,
since it requires $4 \gamma_i g_i = 1/t^* \ll 1$ in contradiction with the initial assumption on $t^*$.
Therefore, we conclude that 

\begin{center}
\it
The bare effective potential analysis does not give a well-defined nor quantitatively approximate result for the definition of the Higgs conformal boundary.
\end{center}

We emphasize that, of course, the definition in \Eref{SCWboundary} goes
way beyond this comparison, as it is valid for any $t^*$, and only $A$ must be
computed in perturbation theory.

\subsection{Conformal Phase Transition}
To complete the picture,
we ask whether $|t^*|$ becomes arbitrarily large when approaching the conformal boundary from the broken side, as naively found from the bare analysis. The implicit question here is whether or not the running couplings flow arbitrarily close to the fixed point value as one approaches the conformal boundary, which is one of the defining signatures of walking dynamics\footnote{Of course, this also implies smooth behavior of the spectrum of the theory as one approaches the boundary. This would be a better definition given that it is scheme independent.}.
It is equivalent to ask the shape of the effective potential at the origin. In fact, this point was already addressed in  section \ref{potential}. We simply need to compute the effective potential at the IR fixed point. We will consider the three possible cases: $A(g^*) > 0$, $A(g^*) < 0$, and $A(g^*) = 0$.
In any practical application, we would need a perturbative IR fixed point, and then it is sufficient to evaluate the improved tree-level coefficient $A^{(0)}(g^*)$ to identify the sign, since perturbative corrections should not become so big to change the sign of $A$ coming from $A^{(0)}$. The delicate case of $A(g^*) = 0$ requires a more involved analysis.  
{ We will, however, keep the discussion general by avoiding referring heavily to perturbation theory.} 
\begin{itemize}
\item[$A(g^*) > 0$] {\it \underline{Jumping with the Parametrically Light Dilaton:}} The effective potential at the origin is convex for any bare values of the coupling constants $g$. This means that if another minimum at a nonzero value of the Higgs field develops the cross-over from the conformal window to the broken side can only be a first order phase transition.  In this case we have always $|t^* | < \infty$ in the broken region implying that simple-minded walking dynamics cannot be realized and no Miransky scaling \cite{Miransky:1984ef,Miransky:1988gk,Miransky:1996pd} emerges, i.e. no RG trajectories in the broken side reach arbitrarily close to the fixed point. In the context of the dilaton and the mass spectrum in the symmetry breaking region, no massive states become arbitrarily light as approaching the conformal boundary due to the potential barrier between the minimum at the origin and the other one. This generates, \emph{de facto}, a mass gap between the two regions. Interestingly, however, the {\it would be} dilaton can still be parametrically lighter than the heaviest states. 
\item[$A(g^*) < 0$] {\it \underline{Running away:}} The effective potential at the origin is concave for any bare values of the coupling constants $g$. Thus either the effective potential always has a non-trivial minimum or is unbounded from below. In any case, the fixed point is unphysical, since in the first case the symmetry will be spontaneously broken and in the second tachyonic states will appear. Thus, there can be no conformal window. All physically meaningful parameter values will lead to spontaneous symmetry breaking and there is no \emph{a priori} reason to expect a dilatonic state to appear. 
\item[$A(g^*) = 0$] {\it \underline{Jumping or Running:}} In this case the conditions of \Eref{diff} and \Eref{SCWboundary} are both non-trivially satisfied for $t^* \to - \infty$ for any initial value of the coupling constants.
The stability of the potential at the origin therefore depends on the specific underlying theory not only at the fixed point but also close to it. 
It can be studied generically in any candidate model by computing the value of $A(g)$ infinitesimally deformed away from the FP. 
This will provide the information on whether $A$ approaches $A(g^*)$ from positive or negative values as the couplings reach the FP. 

\end{itemize}

This ends our generic studies of theories containing scalars with an IR stable FP.
In the following section we will exemplify many of the general features discovered.

 \section{Example of Jumping: QCD with one Adjoint Fermion and Elementary Mesons}
\label{AMS}
It is most instructive to understand the various results above by considering a working example known to feature a Higgs conformal window \cite{Antipin:2011aa}. It is a theory composed of QCD with elementary mesons (Higgs-like) and an extra Weyl fermion transforming in the adjoint representation of the gauge group.
This model has several interesting and topical features. The main focus in \cite{Antipin:2011aa} was to compute the perturbative and non-perturbative infrared spectrum of the model outside of the Higgs conformal window and in particular show that it contains a light dilaton. We will not repeat here  the full theory nor its many features, but encourage the interested reader to become familiar with the model via the basic references  \cite{Antipin:2011aa,Antipin:2012kc}. 

It is here relevant to concentrate on the Higgs potential of the theory which at the Lagrangian level reads:
\ea{
V = \frac{(4 \pi)^2}{N_f^2} \left ( z_1 (\Tr [H ^\dagger H])^2 +z_2 N_f \Tr(H ^\dagger H )^2 \right ) \ ,
}
where $H$ is the Higgs field (mesons) transforming in the bifundamental representation of the chiral symmetry $SU(N_f)_L \times SU(N_f)_R$ and does not transform under the $SU(N_c)$ gauge symmetry. The coupling constants have been appropriately normalized for the Veneziano limit, which is:
\ea{
N_c, N_f \to \infty \qquad \text{while}Ê\qquad  N_f/N_c \equiv x  \quad \text{is kept fixed} \ .
}
This is the limit we will be considering here. The theory features, at the leading order, two perturbative FPs in the beta functions of the model. The value of the couplings at the FPs can be made arbitrarily small by approaching $N_f \to 9N_c/2 = \overline{x} N_c$, where $\overline{N}_f= 9N_c/2$ is the number of flavors where asymptotic freedom is lost. 
 
Explicitly, the two FP values read to first order in $(\bar{x} -x)$:
\ea{
\label{FPs}
z_{1\pm}^* = \frac{-2\sqrt{19} \pm \sqrt{2(8+3\sqrt{19})}}{27}(\overline{x}-x),  \qquad  
z_2^* = \frac{-1+\sqrt{19}}{27}(\overline{x}-x) \ .
}

\begin{figure}[bt]
	\subfloat[Perturbative RG flows around the IR (red) and UV (blue) FPs. The renormalized trajectory connecting the UV and IRFPs is marked as purple.]{\label{RG1}
	\includegraphics[width=0.45\textwidth]{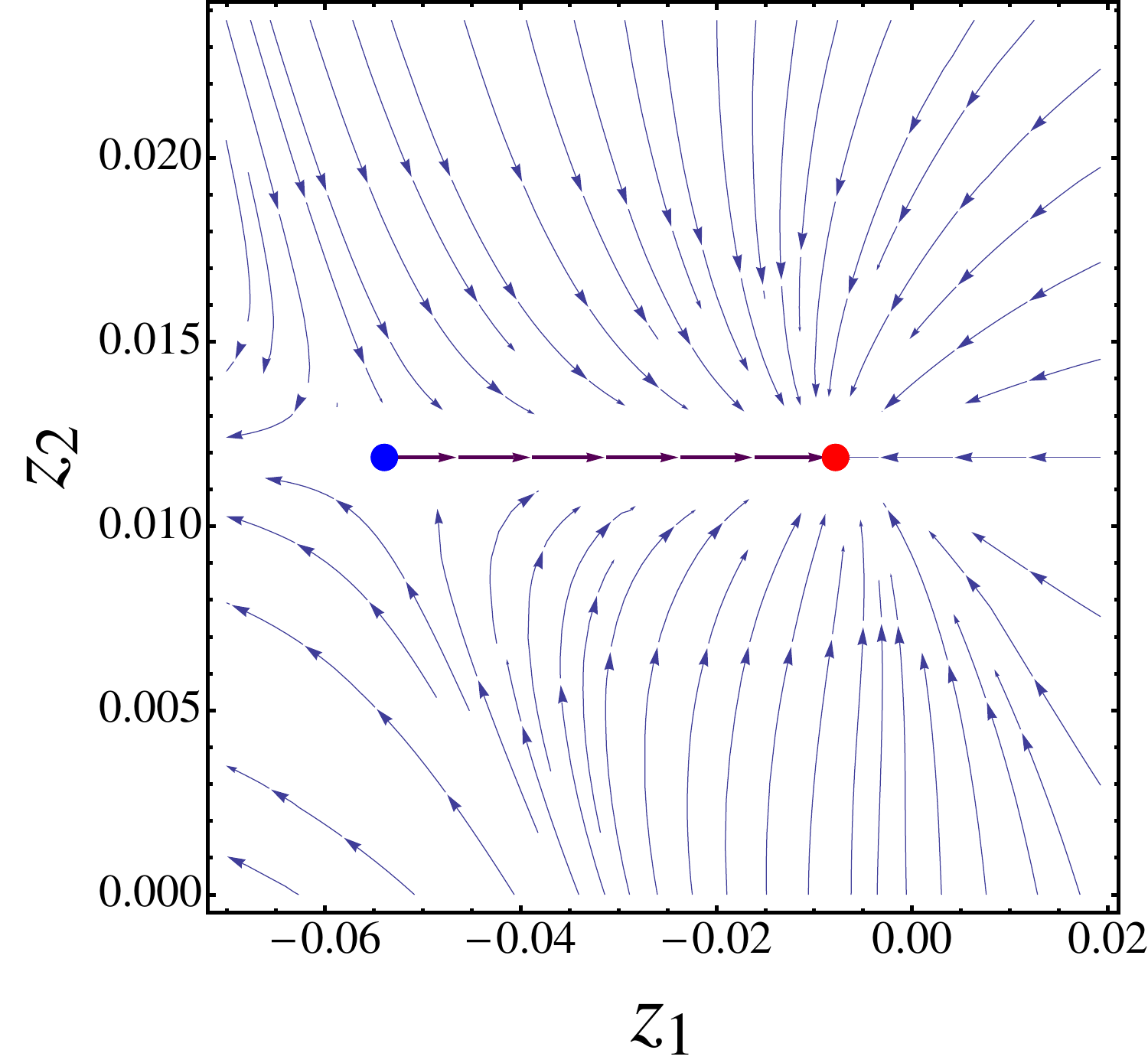}
	}
	\hfill
	\subfloat[The unstable and stable regions against the CW phenomenon. In the unstable region the IRFP is not reached due to phase transition at an intermediate energy scale.]{\label{RG2}%
	\includegraphics[width=0.45\textwidth]{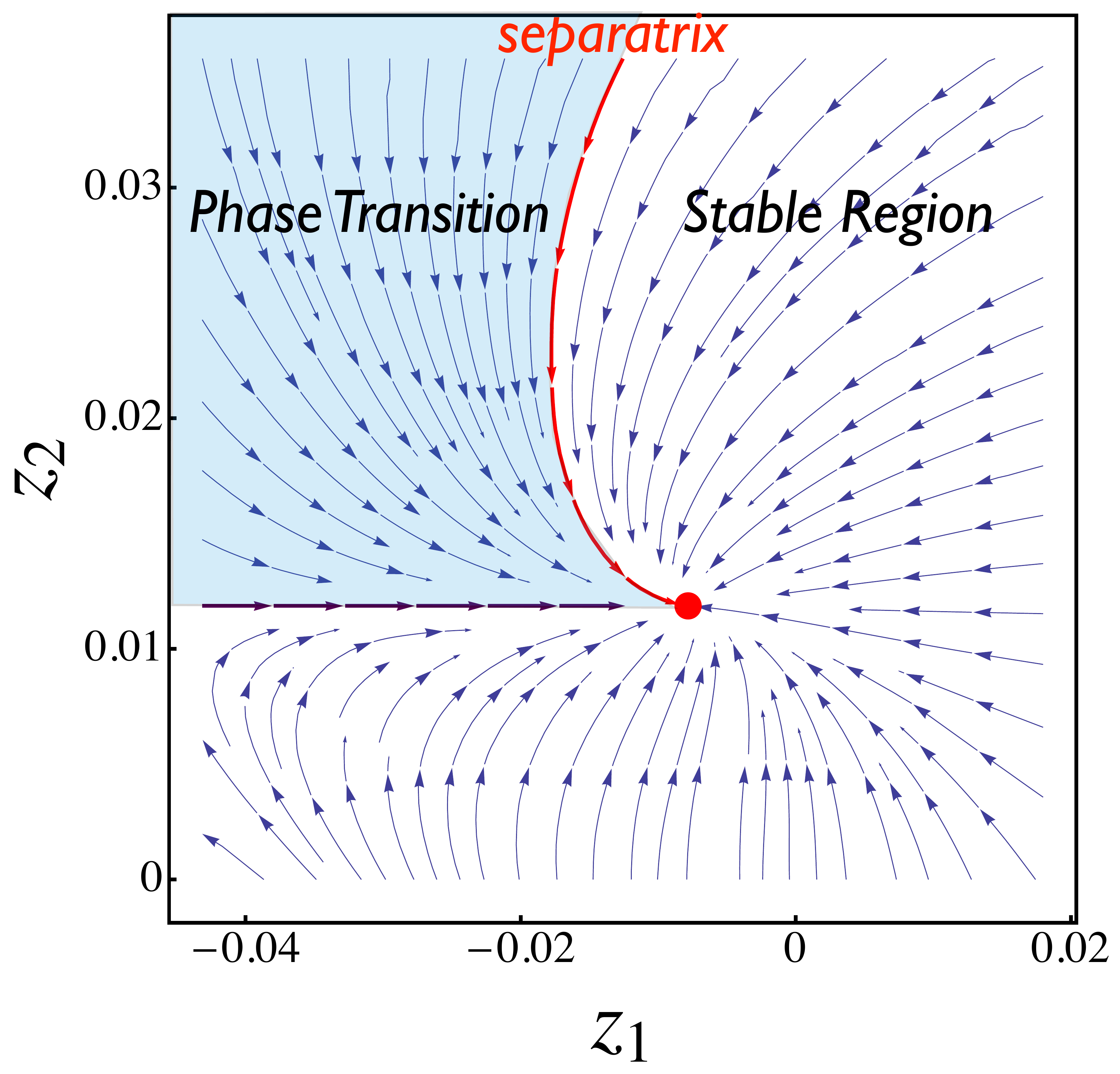}%
	}
	\caption{RG flows in the $(z_1, z_2)$ plane, where it was assumed that the other couplings have quasi-reached the FP \cite{Antipin:2011aa}. For the plots the external parameter was chosen to be $x = 4.41$.}
	\label{RG}
\end{figure}

The FP corresponding to the value $z_{1+}^*$ is the IR stable FP as shown in \cite{Antipin:2011aa}. The other FP corresponding to $z_{1-}^*$ is stable in all directions but one, which is the $z_1$ direction  \cite{Antipin:2011aa}. That means it is an UVFP of the renormalized trajectory connecting the two FPs. An illustration of this is shown in Figure \ref{RG1}.

It was shown in \cite{Antipin:2011aa} that the coupling constant space of this model is partly unstable against the Coleman-Weinberg phenomenon. In this unstable region, the renormalization group (RG) trajectories will not end at the IRFP, due to spontaneous symmetry breaking induced by the quantum corrections to the Higgs potential. This phenomenon is not encoded in the perturbative study of the beta functions. In the stable region, the RG trajectories will flow to the IRFP, and this is the \emph{Higgs conformal window}, as discussed in Section \ref{naiveSCW}. The critical boundary separating the two regions (the separatrix) was determined, as explained, on general grounds in Section \ref{Boundary} and we will also review the determination of the separatrix below. These results are illustrated in Figure \ref{RG2}.

Let us see how it all nicely emerges from the general results of the previous section and what more can be added to this picture. We will also consider and compare with the bare naive analysis.

Consider the classical background field that can break chiral symmetry to the diagonal subgroup\footnote{One can consider this choice to come from simplicity arguments, however, more refined arguments can be found in \cite{Antipin:2011aa,Paterson:1980fc}.}:
\ea{
\label{background}
H_c(x) = \frac{\phi_c(x)}{\sqrt{2N_f}} \mathbf{1} \ .
}
On this background the tree-level potential reads   
\ea{
\widetilde{V} = \frac{V}{\kappa}=A^{(0)} \phi_c^4= (z_1+z_2) \phi_c^4 \ ,
\label{effpot}
}
where we defined a rescaled potential $\widetilde{V}$ with $\kappa = 4\pi^2/N_f^2$ to ease the notation. 
The improved tree-level potential is:
\ea{
\widetilde{V}_{\rm RG}(\phi_c) = A^{(0)}\left(\bar{g}_i(t)\right) \phi_c^4 \ e^{4 \int_0^t \bar{\gamma}\left( \bar{g}_i(s) \right)d s} =  \left [\bar{z}_1(t) + \bar{z}_2(t) \right] \phi_c^4 \ e^{4 \int_0^t \bar{\gamma}\left( \bar{g}_i(s) \right)d s}  \ ,
}
where $\bar{g}_i$ is the set of all running improved coupling constants defined at one loop by
\ea{
\frac{ d \bar{g}_i }{d t} = \frac{\beta(\bar{g}_i)}{1-\gamma(\bar{g}_i)} = \beta(\bar{g}_i) + \mathcal{O}(\bar{g}_i^3)   \ , \quad \text{with} \quad 
\bar{g}(0) = g .
}
In the following we will solve for $\bar{g}_i(t)$ numerically. 

\subsection{ Higgs Conformal Window and then Jump}
The first relevant quantity to compute is the curvature of the effective potential at the origin.
As explained in Section \ref{potential}, this is given by the sign of $A^{(0)}$ at the IRFP, which by using the FP values to leading order in $(\bar{x}-x$) given in Eq.~\eqref{FPs} is:
\ea{
A^{(0)}(g_i^*) = z_{1+}^* + z_2^* \approx 0.042  (\bar{x}-x) > 0  \qquad \forall x<\bar{x} \ .
}
In particular, to one loop order, any point in the coupling constant space of the theory
has a Higgs potential which near the origin reads:
\ea{
\widetilde{V}_{\rm RG} (\delta \phi_c ) \approx   0.042 (\bar{x}-x) \delta\phi_c^4 \ e^{-4 \int_{-\infty}^0 \bar{\gamma}\left( s\right)d s} \ ,
}
for any $\delta \phi_c \ll \mu$. Thus, the effective potential at the origin is a simple convex $\phi^4$ potential. The independence on the bare coupling values is here explicitly seen. 

The boundary between the broken and unbroken phases of the theory is, as described in Section \ref{Boundary}, given by the equation:
\ea{
A^{(0)}(t^*) = \bar{z}_1(t^*) + \bar{z}_2(t^*) = 0 \ ,
}
where $t^*$ satisfies
\ea{
 \frac{dV_1}{d\phi_c} \propto 4 \left [\bar{z}_1(t^*) + \bar{z}_2(t^*) \right] + \beta_1\left(\bar{g}_i(t^*) \right) +\beta_2\left(\bar{g}_i(t^*) \right) &=0  \ ,\label{diffV}\\
 \frac{ d^2V_1(\mean{\phi_c})}{d\phi_c^2} \propto  \left ( 4 + \sum_{\bar{g_i}}
 \beta_{\bar{g}_i} \frac{\partial }{\partial \bar{g}_i} \right ) (\beta_1 + \beta_2) \Big |_{t^*} & > 0 \ . \label{ddiffV}
}
We do not need to know $t^*$ explicitly. In fact, we have the freedom
of renormalization scale invariance to choose $t^* = \ln \mean{\phi_c}/\mu^* = 0$.
Then, since $\bar{g}_i(0) = g_i$, the boundary of the
conformal window is defined by the hypersurface of all RG trajectories
running through the region in coupling constant space satisfying
the conditions:
\ea{
z_1 + z_2 &= 0 \label{zero} \ ,\\
\beta_1(g_i) + \beta_2(g_i) & = 0 \ , \label{zero2}\\
 \sum_{g_j}
 \beta_{g_j}(g_i) \frac{\partial }{\partial g_j} \left(\beta_1(g_i) + \beta_2(g_i)\right ) &> 0 \label{zero3}\ ,
 }
where \eqref{zero} was used to simplify \eqref{diffV} giving \eqref{zero2}, which was then used to simplify \eqref{ddiffV}, giving \eqref{zero3}. This is how the separatrix in Figure \ref{RG2} was found; i.e. it is simply the trajectory which runs through the point where the above conditions are simultaneously satisfied.
 
By this parametrization of the separatrix it follows that 
$\mean{\phi_c}=\mu^*$. {We have already shown that  there is always a minimum at the origin. To guarantee that the conformal phase transition is first order it is therefore sufficient to show that on the separatrix another minimum develops for a nonzero value of $\mu^*$, which is degenerate with the one at the origin. This is indeed the case.  In fact, starting from the very same definition of  $\mu^*$,}
\ea{\label{mus}
z_1 + z_2 = z_1(\mu^*) + z_2(\mu^*) =  0 \ .
}
For any choice of $z_1 + z_2$ at an arbitrary scale $\mu$, the 
couplings will run towards the IRFP. That is,
\ea{
z_1(0) + z_2(0) = z_{1+}^* + z_2^* \approx 0.042 (\bar{x}-x) \neq 0 \quad \forall x < \bar{x}
}
Thus, $\mu^*$ cannot be equal to zero as the condition \eqref{mus} is not satisfied at the FP. Hence, on the separatrix the Higgs potential has
a degenerate minimum with that of the origin at $\mean{\phi_c} = \mu^* \neq 0$

It should now be clear that the phase transition can only be of first order, since i) the potential is convex at the origin and ii) the separatrix is defined where there is a degenerate minimum with that of the origin for $\phi_c = \mean{\phi_c} \neq 0$. The situation, therefore, looks like the one depicted in Figure \ref{SCW}.  By means of a precise computation we will show that this is indeed the case below.

\subsection{Comparison with bare potential}

To really appreciate the importance of the improved analysis, it is best to compare with what one finds from the bare analysis.  On the background \eqref{background}, the one loop bare effective potential is found using e.g. \cite{Martin:2001vx}
\ea{
\widetilde{V} = \frac{V}{\kappa}=\phi_c^4 \left(A+Bt \right) \ ,
\label{effpot}
}
where $\widetilde{V}$ is again rescaled by $\kappa = 4\pi^2/N_f^2$ and where $t\equiv\ln \frac{\phi_c}{N_f\mu}$ is assumed to be $t \lesssim 1$.
The coefficient $A$ takes contributions at both tree level ($0$th order) and one loop ($1$st order), while $B$ arises only at one loop. The explicit one loop values can be found in \cite{Antipin:2011aa} and are not of particular importance here.

In both the bare and improved analysis, we can write the effective potential as a $\phi^4$ potential with an effective running quartic coupling:
\ea{
\widetilde{V}(\phi_c) = \lambda_{\rm eff}(t) \phi_c^4 ,
}
where for each case, $\lambda_{\rm eff}$ is given by
\ea{
\text{Bare:} \quad \lambda_{\rm eff}(t) = A + B t \ , \qquad 
\text{Improved:} \quad \lambda_{\rm eff}(t) =  [\bar{z}_1(t) + \bar{z}_2(t) ]  \ e^{4 \int_0^t \bar{\gamma}\left( \bar{g}_i(s) \right)d s}   
}

\begin{figure}[hbt]
	\subfloat[Bare effective quartic coupling $\lambda_{\rm eff} = 0.01 + Bt$, where from below $B=0.1, 0.05, - 0.001$ .]{\label{BQ}
	\includegraphics[width=0.45\textwidth]{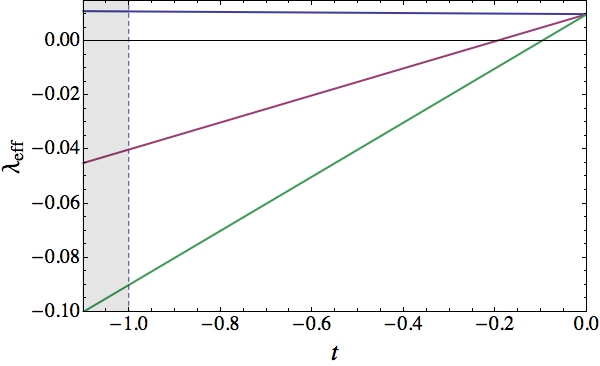}
	}
	\hfill
	\subfloat[Improved effective quartic coupling $\lambda_{\rm eff} = \bar{z}_1(t)+\bar{z}_2(t)$ for three distinct UV initial conditions for the couplings.]{\label{IQ}%
	\includegraphics[width=0.45\textwidth]{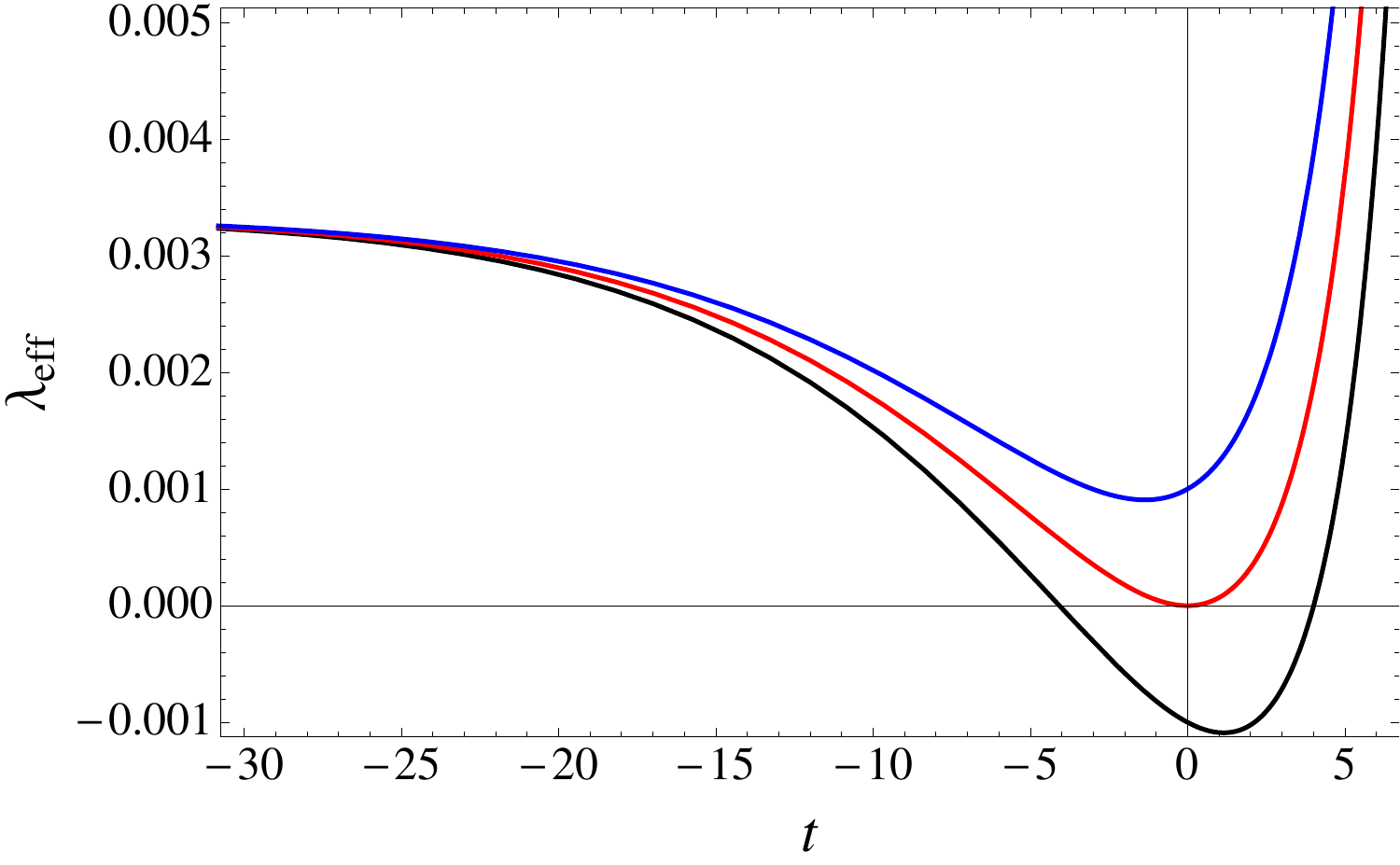}%
	}
	\caption{The effective quartic coupling using the bare and the improved method.  {\it It seems} that for negative $t$ the effective coupling can be negative and therefore the minimum at the origin seems to have vanished. However, the bare analysis cannot be trusted for large values of $t$. The correct physical situation is, in fact, restored by the improved analysis presented in the right panel.  Here the  effective coupling does not change at the origin and remains positive guaranteeing the existence of a minimum at the origin. Furthermore for the three different choices of the UV bare couplings one has, from  top to bottom, either a new local minimum at a nonzero value of $\mu^*$ indicating a new metastable state of the theory (blue line); the phase transition coupling (red line) where the new minimum becomes degenerate with the symmetric one, or finally the case in which the local minimum becomes the actual ground state of the theory (black line). 
Nicely the improved effective coupling stays perturbative in a negatively infinite $t$ range due to the existence of the perturbative IRFP. On the contrary, the bare analysis quickly becomes invalid since both $|t|>1$ and, for very large $|t|$,  $|\lambda_{\rm eff}|>1$.}
	\label{EffectiveQuartic}
\end{figure}

\begin{figure}[!h]
	\subfloat[Bare effective potentials corresponding to the bare effective couplings as given in Fig. \ref{BQ}. The shaded region is where the bare analysis breaks down.]{\label{BP}
	\includegraphics[width=0.45\textwidth]{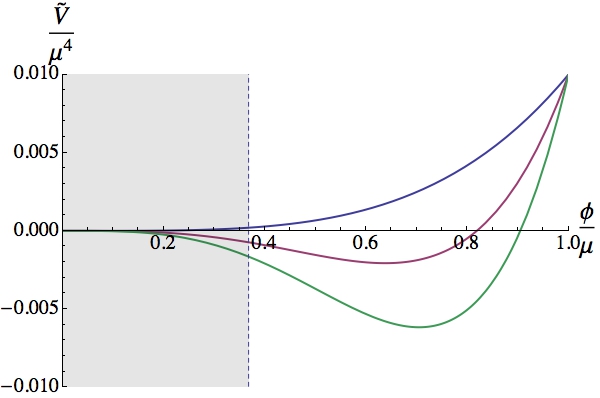}
	}
	\hfill
	\subfloat[Improved effective potential for three different initial conditions, where the upper potential corresponds to one in the conformal window, the middle potential corresponds to theories on the separatrix and the lower potential corresponds to potentials in the spontaneously broken region.]{\label{IP}%
	\includegraphics[width=0.45\textwidth]{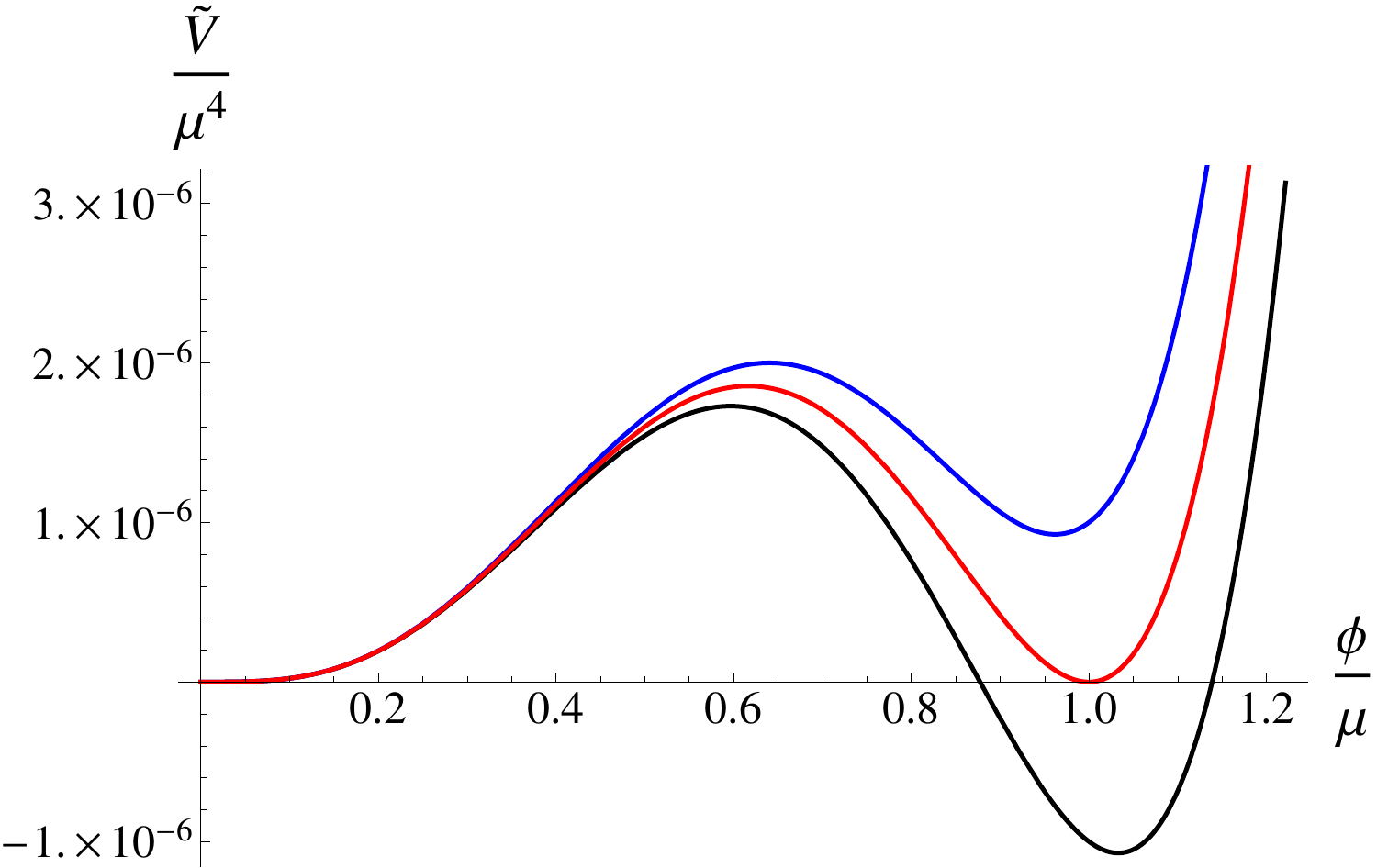}%
	}
	\caption{The effective potential from the two methods. The bare analysis results wrongly in a concave potential at the origin, due to the invalidity of the analysis there (see Figure \ref{EffectiveQuartic}) and leads misleadingly to a smooth phase transition across the separatrix. The improved potential instead reflects the existence and positivity of the IRFP, which shows as expected the potential to be convex at the origin.}
	\label{EffectivePotential}
\end{figure}

The domain of validity in the bare definition is $t \lesssim 1$. In the improved case, we will neglect the exponential term as this will be very close to unity in the perturbative regime we are working and can furthermore not affect the results qualitatively, since it only modifies the scaling property of $\phi_c^4$.
In Figure \ref{EffectiveQuartic} we show
the effective quartic coupling as a function of $t$ in the two methods.
It is seen that the bare analysis does not reflect the existence of the IR fixed, while the improved analysis neatly takes this into account. In fact, from the left panel of Figure \ref{EffectiveQuartic}  {\it it seems} that for negative $t$ the effective coupling can be negative and therefore the minimum at the origin seems to have vanished while a new minimum appeared at a nonzero value of $\mu^*$. However, the point is that  with the bare analysis one cannot trust the result at large and negative values of $t$.  The right panel, in fact, correctly shows that the improved coupling does not change at the origin and remains positive. This implies  that there always is a minimum at the origin. Furthermore for the three different choices of the UV bare couplings one has from top to bottom, either a new local minimum at a nonzero value of $\mu^*$ indicating a new metastable state of the theory (blue line); the phase transition coupling (red line) where the new minimum becomes degenerate with the symmetric one, or finally the case in which the local minimum becomes the actual ground state of the theory (black line). 
We conclude by exhibiting the actual effective potential as found from the two methods and comment on the differences in Figure \ref{EffectivePotential}. The first order nature of the phase transition is manifest. Thus, the model does not exhibit walking dynamics near the conformal boundary, but rather jumping-like dynamics \cite{Sannino:2012wy}.


\subsection{Spectrum and the Light Dilatonic Higgs}

We start with discussing the physical spectrum of the theory in the broken phase. In reference \cite{Antipin:2011aa} we have shown by a direct computation that the classical background field \Eref{background} breaks chiral symmetry to the diagonal subgroup leading to $N_f^2$ Goldstone bosons, $N_f^2-1$ heavy Higgses, 1 pseudo-Goldstone boson associated to the spontaneous breaking of dilatation symmetry, i.e. the dilaton and $N_f\times N_c$ copies of Dirac fermions, transforming according to the fundamental representation of $SU(N_c)$. The fact that the singlet Higgs is the (pseudo) dilaton of the theory relies on reference \cite{Antipin:2011aa} having shown that its mass $m_D$ and decay constant $f_D$ saturate the relation $\langle D | \partial_{\mu} D^{\mu} | 0 \rangle  = - f_D m_D^2  $. In the model presented here $f_D$ is the vacuum expectation value of the scalar field evaluated on the ground state and $m_D$ is the mass of the singlet Higgs state. Furthermore at energies lower than the massive spectrum of the theory a pure ${\cal N}=1$ Super Yang-Mills (SYM) sector emerges composed of the remaining massless underlying fields in the model, i.e. the adjoint fermion (gluino) and the gauge fields (gluons). It was demonstrated that this spectrum further confines and that the associated confining scale is exponentially lighter than the scale of conformal and chiral symmetry breaking  associated to the vev of the scalar field \cite{Antipin:2011aa}. We further argued that the ratio of the dilaton mass to the mass of any other massive states, not associated to the ultra-light SYM sector, vanishes to the improved one-loop order when approaching the separatrix from the symmetry broken phase. The reason for this is that besides the vev there is another control parameter, i.e. a particular combination of the scalar and yukawa couplings, measuring the {\it distance}, in coupling space, from the separatrix. The dilaton mass is proportional not only to the vev but also to this control parameter while the remaining mass spectrum vanishes only because of the vanishing of the vev at the phase transition. We now revisit the discussion about the spectrum of the theory having shown that the Higgs conformal phase transition is not a walking but rather a jumping one. 

The first general observation is that the vev of the theory will jump at the phase transition, i.e. on the separatrix, to the one at the origin of the potential corresponding to the symmetric phase. Therefore we expect the massive spectrum, including the ultra-light SYM composite states, to jump too. To establish the fate of the dilaton, however, we need a more careful understanding of the phase transition properties. One possibility is that  although the vev jumps the control parameter measuring the distance from the separatrix remains continuous. In this event there will still be a very light dilaton in the theory, in principle continuously becoming massless at the phase transition. However the situation can be, in general, more involved as we shall argue below. 

In general the mass spectrum is proportional to the $\mean{\phi_c}\sim \mu$ so that on simple dimensional grounds any mass can be written as  \begin{equation}
 m^2 \sim \mean{\phi_c}^2 f(g_i(\mu))
\end{equation}
where $f(g_i(\mu))$ is a given function of the couplings of the theory. Working consistently to the one-loop order for the beta functions it was shown in \cite{Antipin:2011aa}  that for the $N_f^2-1$ heavy scalars $f(g_i(\mu))\sim z_2(\mu)$, and for the $N_f\times N_c$ fermions, $f(g_i(\mu))\sim a_H(\mu)$ with $a_H\sim y^2$, where $y$ is the Yukawa coupling \footnote{ We manifestly see here that the quartic coupling $z_2$ and Yukawa coupling $y^2$ are to be considered at the same order in perturbation theory as we discussed below \Eref{leadinglog}.}. For the dilaton state $f(g_i(\mu))\sim 4z_2^2(\mu)-xa_H^2(\mu)$ which is the one loop distance from the separatrix. Therefore the dilaton mass vanishes when getting closer to the separatrix \cite{Antipin:2011aa}. Of course, we are using the renormalization group improved approach and therefore the functions $f$ can be consistently approximated by their values assumed at the fixed point. This would seem to imply that there is a light dilaton even for jumping dynamics and it is lighter than the other states of the theory except for the SYM ones. 

There is, however, at least one potential caveat to the picture presented above. It comes from the fact that when the leading-order mass for the dilaton vanishes, higher-order corrections can emerge yielding a nonzero dilaton mass at the phase transition. This, in turn, would mean that the potential on the separatrix near the minimum away from the origin has a positive curvature coming from higher-order corrections. However, even in this case because of the perturbative nature of the model the dilaton will be lighter than the heaviest states of the theory.

\section{Conclusion}
\label{conclusions}
We studied generic properties of the conformal phase transition in gauge theories featuring fermions and scalar particles. The interest in these theories resides in the fact that they are highly topical because of the recent Large Hadron Collider discovery and furthermore allow us to precisely investigate four-dimensional conformal dynamics. We have introduced and characterized the phase transition at the boundary of the Higgs conformal window. 
We provided the necessary conditions
for either a smooth or jumping Higgs conformal phase transition. 
In the perturbative examples known \cite{Grinstein:2011dq, Antipin:2011aa}, only a first order phase transition is, however, realized. 
This implies that the massive spectrum 
of the theory will jump at the phase transition.
Jumping, however, still allows for one of the states, the would-be dilaton of the theory, to be lighter than the heaviest states in the broken phase. Finally we exhibited a calculable Higgs model where, in perturbation theory, we could determine the associated Higgs conformal window as well as the spectrum in the symmetry broken region of Higgs parameter space and could demonstrate that the phase transition is of the jumping type.

\end{document}